\documentclass[]{aa}
\input psfig.tex

\newcommand{\ms}{M$_{\odot}$}
\newcommand{\zs}{Z$_{\odot}$}

\begin{document}


\title{On the early evolution of the Galactic Halo }

\author{     Nikos Prantzos \inst{1} }

\institute{
Institut d'Astrophysique de Paris, 98bis, Bd. Arago, 75014, Paris, France
(prantzos@iap.fr)
}
\date {Received / Accepted }

\abstract
{It is shown that the low metallicity tail of the stellar metallicity distribution 
predicted by simple Outflow models for the Milky Way  halo depends sensitively 
on whether Instantaneous Recycling is adopted or relaxed. 
In both cases, current - and still preliminary - data suggest a 
``G-dwarf problem'' for the halo (reminiscent of the local disk). 
We suggest that the problem  can be solved by introducing
a (physically motivated) early infall phase. We point out several important
implications of such a modification, concerning: the putative Pop. III 
(super)massive stars, the number of stars expected at very low metallicities, the
questions of primary nitrogen and of the 
dispersion in abundance ratios of halo stars. }
\maketitle

\section{Introduction}

The metallicity distribution (MD)
of long-lived stars is one of the most powerful probes of galactic chemical evolution.
The MD of the Galactic halo
field stars (and globular clusters) is quite different from the one of the local disk.
Its peak at a metallicity [Fe/H]$\sim$-1.6\ 
(e.g. Ryan and Norris 1991) points to an {\it effective yield} $y_{HALO}\sim$1/40 \zs \ for Fe.
This has to be compared to the true yield obtained in the solar neighborhood; in the 
framework of a specific infall model - evolution at constant gas amount, which is
a satisfactory approximation to a more realistic treatment  - it can be shown
analytically that the local $y_{DISK}\sim$0.9\zs \ (e.g. Binney and Tremaine 1987).
Taking into account that $\sim$2/3 of solar Fe are produced by SNIa and only $\sim$1/3 by SNII
(e.g. Goswami and Prantzos 2000)
it turns out that the true Fe yield of SNII during the local disk evolution
was $\sim$1/3 \zs \ and,
consequently, the {\it effective Fe yield} of SNII during the halo phase (where they
dominated Fe production) was $y_{HALO}\sim$1/13 $y_{DISK}$. 

The simplest interpretation of such a difference between the
effective yields in the
halo and the local disk remains still the one of Hartwick (1976), who suggested that {\it
outflow} at a rate $F \ = \ k \ SFR$ (where $SFR$ stands for Star Formation Rate) occured
during the halo formation. In the framework of the Simple model of galactic chemical evolution
such outflow reduces the true yield $y_{DISK}$ to its effective value $y_{HALO}=y_{DISK}(1-R)/(1+k-R)$
(e.g. Pagel 1997) where $R$ is the Returned Mass Fraction  ($R\sim$0.33 for a Kroupa et al. (1993)
stellar IMF between 0.1 and 100 \ms); 
the halo data suggest then that the outflow rate was 
$k=(1-R)(y_{DISK}/y_{HALO} -1)\sim$8 times the SFR.
[{\it Note:} This high outflow rate could be interpreted either as gas expulsion after heating
by supernova explosions, or as simple flowing of gas {\it through} the system, e.g. towards 
the Galaxy's bulge].

In this work we focus on the low metallicity tail of the halo MD, a fossile of the earliest 
phase of the halo's evolution.  First, we show that the predictions of the simple outflow
model in the lowest metallicity range ([Fe/H]$<$-3) depend crucially on whether
the Instantaneous Recycling Approximation is adopted or relaxed; to our knoweledge, it is
the first time that this effect is shown to be important in the determination of the
MD of a galactic system. 

The simple outflow model (with or without IRA) 
explains readily the peak and the overall shape of the
halo MD, but it fails to describe the lowest metallicity part of it, as revealed recently
through    the ongoing search of very low metallicity stars by the ``Beers consortium''
(Beers  1999, Norris 1999). The, still preliminary, data suggest that the number of stars
is lower by  factors of $\sim$5-10
than predicted by the simple outflow model at [Fe/H]$\sim$-4; the discrepancy is
larger when the IRA is relaxed (as it should), as we explain in Sec. 2.
This failure gave rise to interesting senarios concerning the physics of the early Galaxy,
such as  the effects of inhomogeneous early evolution
(e.g. Tsujimoto et al. 1999, Oey 2002) or the - often invoked in the past - 
pre-enrichment by a putative population III  of (super-)massive stars
(e.g. Bond 1981, Norris 1999 and references therein).


In this work we suggest a minor - and physically plausible - 
modification to the simple outflow model: as in the solar neighborhood, the
assumption of an early infall phase helps to 
alleviate the discrepancy between theory and observations concerning the very low metallicity
tail of the MD.
In Sec. 2 we discuss the currently available observational data, we explain the differences
between IRA and non-IRA concerning the low metallicity tail of the MD,  and  we show 
how the simple outflow model fails in that metallicity range. We also
present a composite model of early infall + outflow which provides a much 
better fit to the - still preliminary - data. 
In Sec. 3 we discuss the implications of such a senario for the chemical evolution 
of the halo concerning, in particular, the role of a putative Pop. III (super-)massive
stars, the number of stars expected to be found at 
very low metallicities, the timescales of the halo evolution and the dispersion in
element abundance ratios. The forthcoming results of the ``HK survey''
(see Sec. 2) concerning the low metallicity tail of the halo MD, will definitely
allow to check the robustness of the proposed senario.

\begin{figure*}
\psfig{file=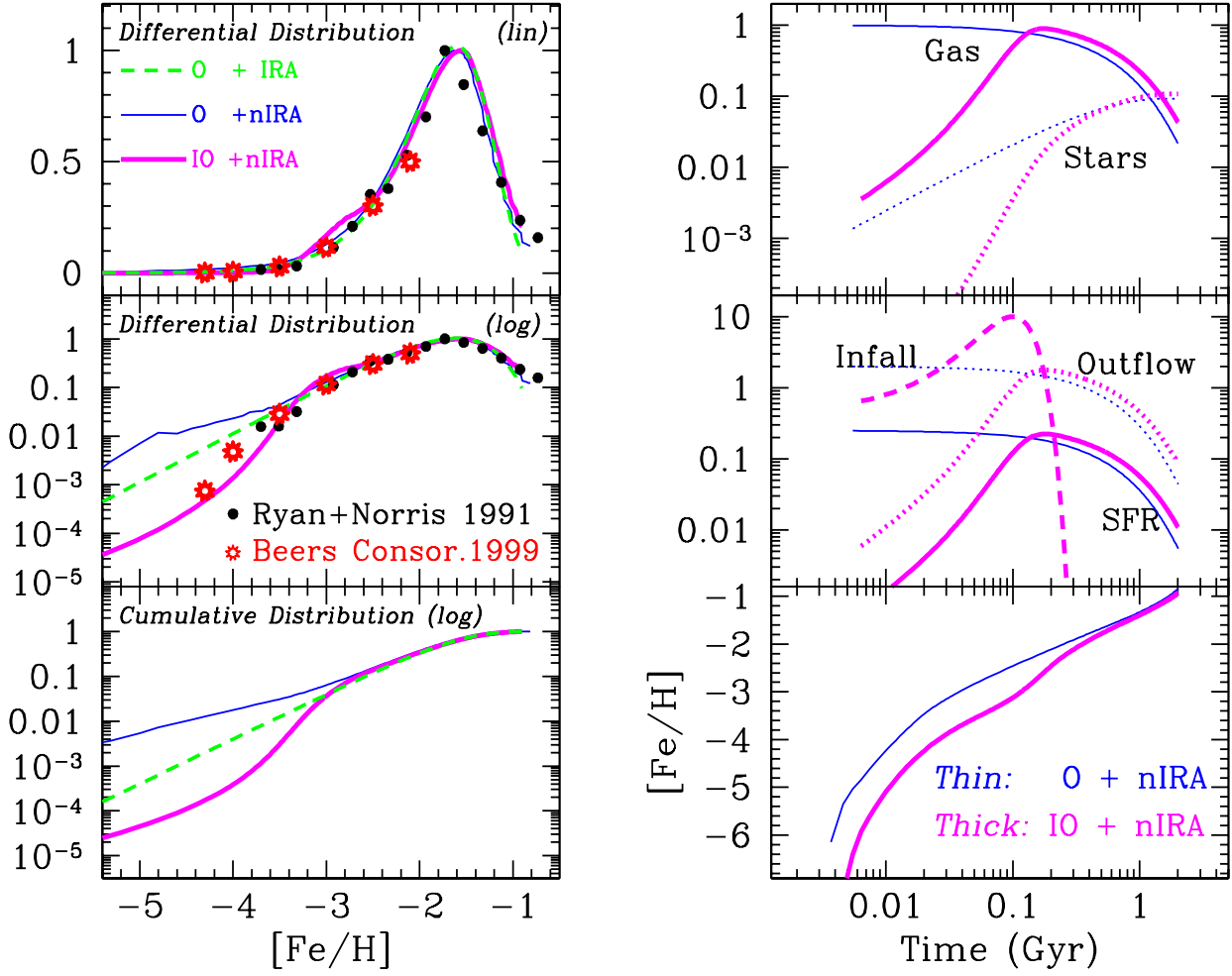,angle=-90,width=\textwidth,height=14cm}
\caption{ Halo metallicity distributions (MD, {\it left}) and features of
relevant models discussed in the text ({\it right}).
{\it Left top:} Differential MD of halo field stars on a linear scale; 
data points from Ryan and Norris
(1991, {\it filled circles}) and the ``Beers consortium'', as presented in Norris  
(1999, {\it asterisks}) ; model results for the pure outflow model with IRA 
(O+IRA, {\it dashed  curves}), the pure outflow model with IRA relaxed 
(O+nIRA, {\it thin solid curves})
and for the early infall + ouflow model (IO+nIRA, {\it thick solid curves}). 
{\it Left middle:} Differential MD of halo field stars on a logarithmic scale;
symbols and curves as previously defined. {\it Left bottom :}  Cumulative 
MD of halo field stars; model curves as previously defined.
{\it Right top:} Evolution of gasesous  mass ({\it solid curves}) 
and stellar mass ({\it dotted curves}) for the pure outflow model with non-IRA ({\it thin curves})
and for the early infall + ouflow model ({\it thick curves}); all quantities are
expressed per unit mass of gas.
{\it Right middle:} Evolution of SFR ({\it solid curves}) and Outflow rate
({\it dotted curves}), as well as of the
infall rate ({\it dashed curve}) for the same  models ; all quantities are
expressed per unit mass of gas and per Gyr.
{\it Right bottom:} Evolution of [Fe/H] in the gas for the same  models.
} 
\end{figure*}

\section{ Construction of the MW halo by early infall}

Two major surveys of the halo field star MD in the 90ies (Ryan and Norris 1991, Carney et al. 1996)
concluded that i) the overall shape of the observed MD is consistent with the predictions
of a simple outflow model {\it a la} Hartwick (1976) and ii) that down to the lowest  metallicity
limit of the surveys ([Fe/H]$\sim$-3) the observed number of metal-poor stars is still
consistent with the predictions of that model; according to Carney et al. (1996): ``... there
is no need to invoke enrichment from  Pop III stars''.

A much more systematic research of extremely metal poor stars is currently performed by T.
Beers and collaborators in the framework of the ``HK survey'', which concerns more
than 10000 metal poor stars. Preliminary results, based on more than half of that sample
have already been presented in Beers (1999) and Norris (1999). Various biases affect
the results  for [Fe/H]$>$-2, while the sample could be considered as representative below
[Fe/H]=-2 (Beers 1999). In the left part of 
Fig. 1 the data of Ryan and Norris (1991) and of the ``Beers consortium'',
i.e. the {\it differential} MD f(Z)= dN/dlogZ (where dlogZ stands for d[Fe/H])
are displayed on a linear scale (upper panel) and on a logarithmic scale (middle panel).
Since the ``HK survey'' data are unaffected by biases only below [Fe/H]=-2, we plot
them only below that metallicity, assuming that at [Fe/H]=-2 they match the corresponding
data of Ryan and Norris (1991). 


Superimposed on the data are the results of three models of galactic chemical evolution.
The {\it dashed curves} correspond to a pure outflow model with IRA and outflow rate
equal to 8 times the SFR, for which an analytical solution can be obtained.
The {\it thin solid curves} correspond to the same model with IRA relaxed 
(numerical solution). In both cases,
the stellar initial mass function (IMF) is from Kroupa et al. (1993) between 
0.1 \ms \ and 100 \ms, the massive star yields from Woosley and Weaver (1995) and
the intermediate star yields from van den Hoek and Groenewegen (1997); both sets
of yields are metallicity dependent. SNIa are included in the calculation,
(see Goswami and Prantzos 2000) 
but their role is negligible in the very early halo phase.
It can be seen  that 

i) The simple outflow model (with or without IRA) fits extremely well the overall
shape of the MD down to [Fe/H]=-3 for the combined sample of Norris and Ryan (1991) and the
``Beers consortium'' (as noticed in many previous studies); 

ii) An increasingly large discrepancy between the IRA and 
non-IRA cases arises below [Fe/H]=-3; this is clearly seen in the middle 
and bottom left panels of Fig. 1, where the differential and cumulative MDs
are plotted, respectively, on logarithmic scales.
This is one of the main results of this work. It can be easily understood as follows:
When IRA is adopted, each stellar generation ejects instantaneously after its formation
the totality of its Fe yield to enrich the intestellar medium. When IRA is relaxed, only
part of that yield becomes available early on (i.e. the Fe ejected by the most massive
and short-lived stars, say first above 50 \ms, then above 30 \ms, then above 20 \ms, etc);
for a constant IMF, this implies that more stars (of all masses)  have to be formed
in that case in order to reach the same metallicity as in the case of IRA.
Of course, after some time, all first generation massive stars have evolved and the yield
reaches its final value; after that moment the two models converge. In Fig. 1 (left middle panel)
that convergence occurs  at [Fe/H]$\sim$-3.5. In the right bottom panel of Fig. 1 it can be
seen that this metallicity is reached after $\sim$35 Myr, i.e. the lifetime of
$\sim$8 \ms \ star, the less massive Fe producers among the massive stars. Obviously, it is the
more ``realistic'' non-IRA  case that should be compared to observations (but see Sec. 4
for a discussion of that point).

iii) An increasingly large discrepancy with the data appears below
[Fe/H]=-3.5 in both IRA and non -IRA cases, but the discrepancy is even more important in 
the non-IRA case:  It reaches a factor of $\sim$12 at the lowest metallicity values  
attained in the HK 
survey. That discrepancy gave rise to interesting discussions on the nature of the first stellar
populations, as mentioned in  Sec. 1.

Despite its spectacular success in the metallicity range 
-3 $<$ [Fe/H] $<$ -1 the pure outflow model
is based on a physically implausible assumption (as is the closed box model), namely that the
system ``waited'' until the total amount of gas became available, and only then started
forming stars and expelling (part of) that gas. Instead, the idea that
star formation started with a low initial amount of gas and the system
continued accreting for some time appears much more plausible (see Sec. 3).
One can see immediately that this idea helps to alleviate the observed paucity of low metallicity
stars, exactly as it helps to cure the G-dwarf problem in the solar neighborhood.

The {\it thick solid curves} on the left panels of Fig. 1 
present the results of a composite model, including
early infall of primordial composition 
(lasting for about 0.2 Gyr) {\it and} outflow as in the simple model.
The form of infall is gaussian (Fig. 1, middle right panel, {\it dashed curves}),
with maximum at 0.1 Gyr and FWHM of 0.04 Gyr and is adopted only for illustration purposes.
It can be seen that this composite model produces a much smaller number of stars 
than the simple outflow model (as expected): by a factor of $\sim$20 (10) at [Fe/H]=-4 and 
$\sim$100 (12) at [Fe/H]=-5, for non-IRA (IRA in parenthesis) models. 
Other forms of infall might lead to even
larger differences with respect to the simple outflow model, but this one matches 
relatively well the currently available - and still preliminary - data.
Gaussian infall has been adopted in the case of the solar neighborhood in Prantzos and Silk
(1998), but despite its two degrees of freedom (maximum value and FWHM) it does not produce
considerably better fits to the data than the simpler exponential infall model (e.g.
Boissier and Prantzos 1999).
 When the final results of the HK survey will become available, it will
be possible to determine the form of the infall rate in the very early 
Galaxy or, more precisely, the early star formation history of the 
halo; this will lead to much tighter constraints on future 
hydrodynamical models of the Milky  Way halo formation 
(see Bekki and Chiba 2001 for a recent attempt).

Inserting the new ingredient of  early infall in the old Outflow ``paradigm'' seems quite 
plausible on physical grounds. Indeed, in any system forming stars on a large scale
(except, perhaps, in starbursts) an early phase of gas accretion 
onto a local gravitational potential well seems unavoidable and during that phase a small
number of stars may be formed.
Outflow due to stellar ``feedback'', i.e. gas heated by kinetic energy of stellar winds and
supernovae, should take place with some time delay, allowing for the collective effects
of many of those stellar sources to operate.
The model presented in Fig. 1 is meant to simulate just such a sequence of events.

At present, it is impossible to derive timescales for the duration of that sequence
of events, based only 
on the observed MD. The situation is different in the solar neighborhood, where
we know: 1) the age of the system to within $\sim$20\% (the age of the local disk
is evaluated to be 8-12 Gyr) and 2) that it reached solar metallicity
4.5 Gyr ago. Combined with the local MD, this information allows us to derive a long
timescale of $\sim$7 Gyr for the formation of the local disk (e.g. Chiappini et al. 1997, 
Boissier and Prantzos 1999). Comparable information is not available for the halo, since the
duration of its formation is known to less than a factor of two (between one Gyr
and 2-3 Gyr).

\section {Implications}

The most important direct implication of the above analysis is that there is no need
to invoke a Pop III of zero metal (super-)massive stars in order to explain the 
observed halo MD; in a similar way, the infall hypothesis makes obsolete the need
of a local disk pre-enrichment by the halo stars in order to solve the G-dwarf problem.
Of course, the second chemical observable of halo stars, namely abundance ratios with
strange behaviour at the lowest metallicities (like Mn/Fe or Cr/Fe, e.g. McWilliam 1997), could 
still be invoked to justify the existence of Pop III stars (e.g. Umeda and Nomoto 2002).
However, current uncertainties in the mecanism of core collapse supernova
explosion (e.g. Janka et al. 2002) do not allow us to draw any firm conclusion on the
expected yield ratios from those objects  at any metallicity (see Chieffi and Limongi 2002
for a recent analysis of yields of stars of zero initial metallicity). 
Therefore, to the author's
opinion, there is currently no {\it strong} observational argument for the existence 
of (super-)massive Pop III  stars (or of a very peculiar stellar IMF) preceding the
formation of the galactic halo and affecting its
early chemical evolution.

A second implication, closely related to the first one, concerns the number of 
stars expected to be detected at very low metallicities by current or future surveys. 
The early infall + outflow model presented in Fig. 1 has a fraction of 3 10$^{-4}$ 
stars with [Fe/H] $<$-4 and 3 10$^{-5}$  stars with [Fe/H] $<$-5 (Bottom left panel).
These numbers suggest that the current record holder of low Z stars, namely the giant
HE 0107-5240 of [Fe/H]=-5.2 (Christlieb et al. 2002) may well remain the only of its kind
not only at the completion of the HK survey ($\sim$10 000 stars) but also
in future surveys exploring samples ten times larger.
The quest for objects with even lower metallicities will require million-object samples.

Taking into account that the mass of the halo is estimated to $\sim$10$^9$ \ms 
(Morrison 1993),
that the Kroupa et al. (1993) IMF has $\sim$2 stars per unit mass below 0.8 \ms,  
and assuming that the aforementioned 
infall + outflow model applies to the halo as a whole, one sees 
that there must be about 5 10$^5$ stars with  [Fe/H]$<$-4, 5 10$^4$ stars with 
[Fe/H]$<$-5, and even a few thousand stars  with [Fe/H]$<$-6 (note that for a normal IMF 
extending down to 0.1 \ms, the vast majority of them are red dwarfs). Clearly, {\it if}
(and that is a big {\it if}) the proposed modification to the simple  outflow model
describes satisfactorily the early halo evolution, there is a great potential
for discoveries of large numbers of extremely low Z stars by future (very)
large scale surveys, like the GAIA mission (see de Zeeuw 2002 for a summary of the 
perspectives of GAIA).

A third implication of the above results concerns the timescales of the early halo
evolution. In Fig. 1 (Bottom right panel) it can be seen that in the case of the simple
outflow model it takes only 35 Myr to reach [Fe/H]=-3, while it takes almost
three times as long ($\sim$100 Myr) in the case of the infall + outflow model.
In the former case, only massive stars have time to enrich the interstellar medium;
in the latter, even stars with M$>$ 4 \ms \ have time to evolve up to the AGB phase
and eject their products of shell He-burning (C and s-process elements) 
or of Hot Bottom Burning (HBB), i.e. primary nitrogen.

The answers on several important current problems of galactic chemical evolution
may depend on those timescales. For instance, the N/Fe ratio in the Milky Way seems to
be constant and $\sim$solar as a function of [Fe/H], from [Fe/H]$\sim$-3 up to [Fe/H]=0.
This constancy suggests that N is produced as a primary element (i.e. with a 
yield independent on the initial metallicity of the star) quite early in the Milky Way.
In the pure outflow model, only massive stars can be invoked for such an early production
of primary nitrogen,
but all current models of massive star nucleosynthesis produce only secondary N;
even the recent massive star models  including rotation 
(Meynet and Maeder 2002)  produce insufficient amounts of primary N 
to explain the observations (Prantzos 2003). 
In the infall + outflow model, intermediate mass stars have time enough to 
evolve before the metallicity reaches
[Fe/H]=-3 and could naturally produce
the required primary nitrogen, either by HBB or by rotational mixing.

A fourth implication of the different timescales obtained in the simple and modified
halo models concerns the dispersion  in the elemental abundance ratios at low metallicities.
From the theoretical point of view, it is expected that mixing of stellar ejecta with 
the ISM should have been incomplete in the early Galaxy, when the mixing timescale
of a few Myr was comparable to the evolutionary timescale of the system and of the first
stellar generation (Audouze and Silk 1995, Tsujimoto et al. 1999,  Argast et al. 2000, 
Ishimaru et al. 2003). However, Carretta et al. (2002) find that, down to the lowest
metallicity reached by current surveys ([Fe/H]$\sim$-3), dispersion in abundance ratios
of intermediate mass elements is no larger than expected from observational uncertainties.
The question ``{\it When} was the metallicity [Fe/H]=-3 reached during the halo evolution?''
becomes obviously important in that context and  the answer may depend (also) on the
modifications of the simple model discussed here. 
Note, however, that in the framework of inhomogeneous chemical evolution models
Tsujimoto et al. (1999) and   Argast et al. (2000) obtain ``naturally''
much larger timescales than in all the models discussed in Sec. 2.

\section {Summary}

In this work it is shown that the low metallicity tail of the
MD predicted by simple Outflow models for the MW halo depends sensitively
on whether IRA is adopted or relaxed. 
In both case, current - and still preliminary - data suggest a 
``G-dwarf problem'' for the halo. We suggest that it can be solved by introducing
a physically motivated early infall phase; we adopt here a gaussian infall rate, 
which turns out to fit the available data well, but 
determination of the  precise
form of the infall will have to wait for the results of complete
surveys in the future. We point out several important
implications of such a modification, concerning: the putative Pop. III 
(super)massive stars, the number of stars expected at very low metallicities, the
questions of primary nitrogen and of the 
dispersion in abundance ratios of halo stars.
The forthcoming results of the ``HK survey''
concerning the low metallicity tail of the halo MD, will definitely
allow to check the robustness of the proposed senario.

The phenomenological analysis of this work does not take into account the modern
framework of early  galaxy formation in a cosmological context. First results of
such studies have been recently presented (Bekki and Chiba 2001) and the issue of
the halo MD has been successfully addressed, albeit without a clear physical
explanation for that success. 

\acknowledgements
{I am grateful to the referee, U. Fritze von Alvesloeben, for her constructive
criticism.}

{}

\end{document}